\documentclass[letterpaper,prd,nofootinbib,floats,superscriptaddress,eqsecnum,tightenlines, showpacs,11pt]{revtex4}
    
\usepackage{enumerate}
\usepackage{graphicx}
\usepackage{subfigure}
\usepackage{showlabels}

\usepackage{array}
\usepackage{setspace}

\usepackage{hyperref}

\usepackage{color}

\usepackage{amsmath,amssymb,amsfonts,amsthm,latexsym,stmaryrd}

\newcommand{\be}{\begin{equation}}
\newcommand{\ee}{\end{equation}}
\newcommand{\beq}{\begin{eqnarray}}
\newcommand{\eeq}{\end{eqnarray}}
\newcommand{\bea}{\begin{eqnarray}}
\newcommand{\eea}{\end{eqnarray}}
\newcommand{\beqn}{\begin{eqnarray}}
\newcommand{\eeqn}{\end{eqnarray}}

\newcommand{\lbr}{\llbracket}
\newcommand{\rbr}{\rrbracket}

\def\ext{\mathrm{ext}}

\def\L{L}
\def\I{I}

\def\rd{\mathrm{d}}
\def\pa{\partial}

\def\co{\chi}
\def\hco{{\hat\chi}}

\def\hxi{{\hat{\xi}}}
\def\cL{{\cal L}}
\def\curv{{\cal F}}

\def\ip{\lrcorner\,}
\def\curv{{\cal R}}

\def\h={\hat{=}}
\def\cF{{\cal F}}
\def\cS{{\cal S}}

\newcommand{\Diff}{\text{Diff}}
\newcommand{\R}{\mathbb{R}}

\begin{document}

%\title{Translation symmetry in gravity}
%\title{Gauge versus symmetry in gravity}
%\title{Non-asymptotic symmetries of gravity}
\title{
A canonical bracket for open gravitational system}
%
%\author{William Donnelly}
%\email{donnelly@physics.ucsb.edu}
%\affiliation{Department of Physics, University of California, Santa Barbara \\
%Santa Barbara, California 93106, USA}

\author{Laurent Freidel}
\email{lfreidel@perimeterinstitute.ca}
\affiliation{Perimeter Institute for Theoretical Physics\\
31 Caroline St. N, N2L 2Y5, Waterloo ON, Canada}

\date{\today}

\begin{abstract}
This paper shows that the generalization of the Barnich-Troessaert bracket recently proposed to represent the extended corner algebra can be obtained as the canonical bracket for an extended gravitational Lagrangian. This extension effectively allows one to reabsorb the symplectic flux into the dressing of the Lagrangian by an embedding field. It also implies that the canonical Poisson bracket of charges, forms a representation of the extended corner symmetry algebra.
\end{abstract}

\maketitle

\newpage
\thispagestyle{empty}
%\tableofcontents
\newpage

\section{Introduction}

One of the most vexing questions of theoretical physics is whether the formation and subsequent evaporation of black holes is a unitary process, and if so, how the information comes out \cite{Hawking:1975vcx,Giddings:2020zso,Jacobson:2019gnm}.
Answering this question requires knowing how much information can be encoded in a black hole, and how it is transferred to the exterior. In addition, resolving this problem requires understanding what are the quantum black-holes hair \cite{Hawking:2016msc} and their semi-classical analog?

This question is, in fact, related to a more fundamental challenge which is 
 to describe the phase space of a compact region in general relativity.
This problem is more basic than questions about black holes: it underlies the fundamental puzzle of how one can define local subsystems in a theory of gravity \cite{Donnelly:2016auv, Donnelly:2016rvo, Donnelly:2017jcd}  which are necessary to describe experiments performed by localized observers inside the system.

This second question has received renewed attention in recent years. The series of work 
\cite{Donnelly:2016auv, Speranza:2017gxd, Freidel:2020xyx,Freidel:2020svx,Freidel:2020ayo,Ciambelli:2021vnn,Freidel:2021cjp} has established that there is a universal symmetry group called the extended corner symmetry group,  associated with the corners of the spacetime causal diamonds and representing gravitational subsystems.  
It has been established that the extended corner symmetry group acts on the gravitational phase space associated with finite regions and that symmetry charges for this group represent the hair needed to reconstruct the gluing of subregions
\cite{Rovelli:2013fga,Donnelly:2016auv}. Moreover, this extended corner symmetry group has been shown to be a maximal \cite{Ciambelli:2021vnn} and universal \cite{Speranza:2017gxd} subgroup of space-time diffeomorphisms.

In \cite{Donnelly:2016auv} a subgroup called the corner symmetry group was identified and shown to be represented canonically on the gravity phase space.  In other words, the Noether charges for the corner symmetry action are Hamiltonians that implement the symmetry through the canonical bracket. This part of the symmetry is readily quantizable. The study of the corner symmetry group representations has been initiated in \cite{Donnelly:2020xgu}, where complete sets of Casimir have been identified. The rest of the extended corner symmetry group includes the normal supertranslations, which have not yet received a satisfactory canonical interpretation.

The challenge, in this case, is formidable.  The supertranslations are spacetime transformations that translate the corners. Such transformations do not preserve the gravitational symplectic structure, and therefore, they fail to be Hamiltonian. The leak of symplectic flux through the corner is encoded into a variational one form that represents the mathematical obstruction to have integrable supertranslation charges.  At the physical level, and on the asymptotically flat phase space, the non vanishing of flux expresses the presence of gravitational radiation \cite{Ashtekar:1981bq,Dray:1984rfa,Ashtekar:1990gc, Wald:1999wa,   Barnich:2009se,Barnich:2010eb, Barnich:2011mi, Barnich:2013axa,Compere:2018ylh, Compere:2020lrt,Freidel:2021yqe, Freidel:2021qpz,Chandrasekaran:2021vyu}.
Of course, gravitational radiation prevents the existence of isolated gravitational subsystems.
 Therefore constructing an Hamiltonian action of  supertranslations is tantamount to understand if one can represent open systems canonically. Presented in this way, this sounds like an impossible task since one usually expects that open systems do not lend themselves to quantization. Therefore what is usually done is to impose by hand boundary conditions that close the system as in AdS/CFT  
\cite{Henneaux:1985tv,Skenderis:2002wp,Marolf:2008mf,Andrade:2011dg} or study topological field theories and reabsorb the Hamiltonian flux into a field redefinition \cite{Ruzziconi:2020wrb,Geiller:2021vpg}. Unitarity and canonical representation of the charges is achieved in these cases because gravitational radiation is either conveniently killed or dynamically absent.
If one wants to understand non perturbatively the nature of quantum radiation, one must allow symplectic flux to be non-vanishing.  In recent years, several studies of open Hamiltonian systems associated with null boundaries with non-zero flux have advanced our knowledge 
of fluxes along null surfaces \cite{Donnay:2015abr,Donnay:2016ejv,Donnay:2019jiz,Hopfmuller:2016scf,Hopfmuller:2018fni,Adami:2020amw, Grumiller:2019ygj, Grumiller:2020vvv,Chandrasekaran:2018aop, Chandrasekaran:2020wwn,Adami:2021nnf}

Recently, two works have addressed the canonical representation of the extended corner symmetry group. In \cite{Freidel:2021cjp} a generalization of the Barnich-Troessaert charge bracket \cite{Barnich:2011mi} has been constructed. It was shown that, under this bracket, the charges of corner symmetry form a representation of the corner symmetry algebroid.  The problem of interpreting this bracket in terms of a Poisson bracket on a phase space was left open.
Besides, the work \cite{Ciambelli:2021vnn} postulated a canonical representation of supertranslations on a phase space that includes the embedding field introduced in \cite{Donnelly:2016auv}.  However, this construction was not connected to a symplectic structure derived from a Lagrangian, and the relation with the generalized  Barnich-Troessaert was left unresolved.

In this note, we provide a  resolution of these issues. We demonstrate that the dressing of the Lagrangian by an embedding field provides a modification of the symplectic form, which allows a canonical representation of the supertranslations. We also establish that this modification of the symplectic structure gives the generalized Barnich-Troessaert bracket. This results therefore identify the two strategies of \cite{Ciambelli:2021vnn}  and \cite{Freidel:2021cjp} as equivalent.

During the completion of this project, we became aware of the independent work by Ciambelli, Leigh and Pai \cite{ciambelli2021embeddings} reaching a result similar to ours.

We begin in section \ref{CoPS} by giving a self-contained derivation of the construction of the diffeomorphism Noether charge for a covariant Hamiltonian. This section summarizes the main results of \cite{Freidel:2021cjp}. We also present the construction of the 
the pseudo-bracket of charges.
In section \ref{EPS} we introduce the embedding field and the Lagrangian's dressing. We show that this dressing can be reabsorbed in terms of the variational calculus by a field space connection. We then prove that the resulting extended symplectic potential gives after contraction with the diffeomorphism generators the generalized Barnich-Troessaert charge bracket. It forms a representation of the extended corner symmetry algebra.

\section{Covariant Phase space}\label{CoPS}

The construction of the covariant phase space requires the introduction of a bicovariant calculus \cite{Anderson, Compere:2018aar, Barnich:2001jy}which is a generalization of Cartan calculus including both spacetime Cartan derivatives and  variational field derivatives. We follow here the exposition of \cite{Freidel:2020xyx, Chandrasekaran:2020wwn, Freidel:2021cjp}.  We assume that our fields  denoted $\phi^A$ are smooth section of a Bundle $\pi: P \to M$ over spacetime $M$ and we denote by $\mathbb{F}$ the space of fields.
The  automorphisms group $\mathrm{Aut}(P)$ of the bundle is generated by the fiber-preserving diffeomorphisms:  A diffeomorphism $\Psi:P\to P$
 is called fiber-preserving if $\pi \circ \Psi =\psi \circ \pi$
 for some smooth diffeomorphism $\psi: M\to M$ of $M$.  If $\psi=\mathrm{id}_M$ then 
 $\Psi$ is  a gauge transformation. At the infinitesimal level an automorphism is represented by a vector field $\xi \in \mathfrak{X}(P)$ such that $\bar\xi \circ \rd \pi = \rd \pi\circ \xi $,
 where $\rd \pi: TP \to TM $ is the anchoring map, and $\bar{\xi}\in \mathfrak{X}(M)$ is a vector field on spacetime. 
 When $P$ is  a G-bundle associated the group action  $G\times P \to P$ the automorphism has a semi-direct product structure $\mathrm{Diff}(M) \ltimes G^S$.

\subsection{Bicovariant calculus}

Given a vector field $\xi$ representing an infinitesimal automorphism, we denote by $\iota_\xi$ the interior product.
We also denote $\rd $ the Cartan differential. Both $\rd $ and $\iota_\xi$ are graded differentials acting on spacetime forms of respective weight $+1$  and $-1$. All commutators that follow are taken as graded commutators.
The Lie derivative along $\xi$ is defined as the commutator  $\cL_\xi=[\rd, \iota_\xi]$.
The map \be
\xi \to \hxi:= \cL_\xi,
\ee  denotes the lift of $\xi$  onto field space.  This means that  the action of the field space Lie derivative on the set of fundamental fields $\phi^A$ is given by $L_{\hat{\xi}}=\cL_\xi \phi^A$. To fix the idea we can assume that the space of fields contains the metric $g_{ab}$ and some scalar fields $\phi$. In this case the automorphism group is simply $\mathrm{Diff}(M)$ and $\xi$ is a vector field on $M$ with action 
$L_{\hat{\xi}} g_{ab} =\nabla_a\xi_b+\nabla_b\xi_a$ and  $L_{\hat{\xi}}\phi = \xi^a\pa_a \phi.$
We also have a Cartan calculus on field space where $\delta $ is the field space variational differential, and we denote by $\I_{\hxi} = {\cal L}_\xi \ip $ the field contraction  and by 
$\L_\hxi=[ \I_\hxi, \delta ]$, the field space Lie derivative along $\hat{\xi}=\cL_{\xi}$.
The variational Cartan calculus axioms imply that 
\be
[I_{\hxi}, I_{\hat\psi}]=0, \qquad
[L_\hxi, \delta ] =0,\qquad
[L_\hxi, I_{\hat\psi}]= - I_{\widehat{\lbr \xi , \psi \rbr}},\qquad
[L_\hxi, L_{\hat\psi}]= - L_{\widehat{\lbr \xi , \psi \rbr}},
\ee
where the bracket is an algebroid bracket  generalizing of the Lie bracket for field dependent vector fields,\footnote{ The algebroid  bracket is evaluated by using the definition $L_\hxi \phi = -\cL_{\xi}\phi$, when acting on a fundamental field, and evaluating
\bea
[L_\hxi, L_{\hat\psi}]\phi &=&  [L_\hxi \cL_{\psi} \phi - L_{\hat\psi} \cL_{\xi} \phi]
=   \cL_{L_\hxi \psi} \phi  + [\cL_{\psi}, \cL_\xi] - \cL_{ L_{\hat\psi} \xi} \phi  
= - \cL_{\lbr \xi , \psi\rbr} \phi = -L_{\widehat{\lbr \xi , \psi\rbr}}\phi.
\eea
} i-e vector field such that $\delta \xi\neq0$:
\be\label{Lie-mod}
  \lbr\xi,\psi \rbr:= {[\xi, \psi]_{\mathrm{Lie}} +  L_{\hat\psi} \xi}-L_{\hxi} \psi \,.
 \ee
 A general field observable is a field space form $O(\phi^A,\delta \phi^A)$ that can be expressed entirely in terms of the physical fields and their variations.
 We say that $O(\phi^A,\delta \phi^A)$ is a covariant observable when 
\be
\L_{\hat\xi} O= \cL_\xi O+ \I_{\delta \xi} O.
\ee
When the vector fields are field independent this simply means that 
$(\L_\xi-\cL_\xi)O=0$. In this case
we can replace the action of the field space Lie derivative with the infinitesimal automorphism group action. 
Examples of covariant covariant observables includes the fundamental fields and their variational differential. Indeed we have that $\L_\xi \phi^A =\I_\hxi \delta \phi^A= \cL_\xi\phi^A$ by definition.
While on the  basis of one form we get 
\be
\L_\xi \delta \phi^A = \delta \I_\hxi \delta \phi^A= \delta \cL_\xi\phi^A=
\cL_\xi \delta\phi^A + \cL_{\delta \xi}\phi^A  = (\cL_\xi + \I_{\delta \xi} )\delta \phi^A.
\ee
This leads to the important concept of the anomaly \cite{Hopfmuller:2018fni, Chandrasekaran:2020wwn, Freidel:2021cjp} of an observable which is given by
\be
\Delta_\xi O :=(\L_\xi - \cL_\xi - \I_{\delta \xi})O.
\ee
\subsection{Covariant Phase space and fluxes}\label{CovPS}
Given a Lagrangian $L$,  we can construct from it, using Anderson's homotopy \cite{ Anderson} operators, a unique \cite{LeeWald, Freidel:2020xyx, Freidel:2021cjp} symplectic potential $\theta$  and equation of motion $E$ satisfying the constraints
\be
\delta L = \rd \theta - E.
\ee
Taking the field-space differential of this equation gives the conservation equation 
\be\label{cons}
\rd\omega =\delta E, 
\ee
where $\omega:=\delta \theta$ is the symplectic form density.
The conservation equation for the symplectic form is the classical equivalent of the unitarity condition.

A symmetry generator 
$\xi \in \mathrm{Aut}(P)$ is an element of the automorphism group of the bundle.
By definition a field transformation $\hxi$ is a Lagrangian symmetry
when $\hxi[L]= \rd \ell_\xi$. We are only interested in local symmetries which are such that 
\be\label{bianchi}
I_\hxi E = \rd C_\xi,
\ee
where $C_\xi$ is the constraint which vanishes when $E$ vanishes.\footnote{For Einstein gravity we have that $C_\xi = \xi^\mu (G_\mu{}^\nu-8\pi G T_\mu{}^\nu)\epsilon_\nu$. 
Where $G_\mu{}^\nu$ is the Einstein tensor, $T_\mu{}^\nu$ is the energy-momentum tensor and $\epsilon_\mu= i_{\partial_\mu} \epsilon $ is the codimension $1$ volume form.
}
The Lagrangian is said to possess no anomalies when $\Delta_\xi L=0$, which implies $\ell_\xi=\iota_\xi L$. The construction of the  symplectic potential is covariant which means that we also have $\Delta_\xi\theta=0$ in this case. We restrict our analysis to covariant Lagrangians. Readers interested in the generalization of the 
phase space analysis to non-covariant Lagrangian are referred to \cite{Freidel:2021cjp}. For covariant Lagrangians we have that  \cite{Iyer:1994ys, Wald:1999wa, Barnich:2001jy, Harlow:2019yfa}
%\bea
%\L_{\hxi} L =  I_\hxi \delta L = \rd I_\hxi \theta - I_\hxi E= \rd (I_\hxi \theta - C_\xi),
% \qquad \cL_\xi L = \rd (\iota_\xi L ).
%\eea
%Taking the difference we find that 
\be
\Delta_\xi L
%= I_\hxi \delta L - \rd \iota_\xi L 
= \rd( I_\hxi \theta -\iota_\xi L - C_\xi)=0. 
\ee
This means that the Noether current is given by  
\be\label{Ncur}
 J_\xi:= \I_\hxi \theta -\iota_\xi L = C_\xi + \rd q_\xi.
\ee
The Noether current is therefore the sum of a constraint $C_\xi$ that vanishes on-shell and of a corner charge aspect\footnote{For Einstein gravity minimally coupled to matter we have that  $q_\xi$ is given by the Komar expression $q_\xi = \frac1{16\pi G}  \star \rd g  (\xi) =\frac1{16\pi G}  \epsilon_{ab} \nabla^a\xi^b$ where $\epsilon_{ab}=\iota_{\pa_a}\iota_{\pa_b} \epsilon$ is the codimension $2$ volume form.} $q_\xi$.
 The current  conservation equation $ \rd J_\xi =I_\hxi E$ is trivial since 
 the current is exact on-shell.
 
 The main theorem of Noether stems from the fact that 
the field transformations are Hamiltonian generators. 
In practice this means that we have the following fundamental canonical relation for field dependent symmetries
\be \label{Fundr}
-\I_\hxi \omega = \underbrace{\delta  \left(  C_\xi + \rd q_\xi  \right)}_{\mathrm{Noether\, Charge}} -[C_{\delta \xi} + \rd q_{\delta \xi}] - \underbrace{\left[\iota_\xi E+ 
\rd (\iota_\xi \theta) \right]}_{\mathrm{Symplectic\, Flux}}.
\ee
A simple proof is provided in appendix \ref{proofA}. If one integrates this on a slice $\Sigma$, with boundary $S=\pa \Sigma$,  and defines $\Omega =\int_\Sigma \omega$, the fundamental Noether theorem reads 
\be\label{canonical}
-\I_\hxi \Omega =  \delta Q_\xi - Q_{\delta\xi} -  \cS_\xi,
\ee
where $Q_\xi$, the Noether charge associated with the Lagrangian $L$, is  given by 
\bea 
Q_\xi :=  \int_\Sigma C_\xi + \int_{S} q_\xi. 
\eea
The charge is a boundary term when the constraints are satisfied.
 This charge allows us to distinguish the diffeomorphisms which are gauge transformations form the ones which are symmetries. By definition the gauge transformations are represented by vanishing on-shell Noether charges while the symmetries are associated with transformations that carry non-vanishing on-shell Noether charges. The subgroup of diffeomorphisms which represents a symmetry of a covariant Lagrangian for a slice $\Sigma$ with boundary $S$ is the extended corner symmetry group given by \cite{ Donnelly:2016auv, Ciambelli:2021vnn, Freidel:2021cjp} 
 \be
H_S:=\left( \mathrm{Diff}(S)\ltimes \mathrm{GL}(2,\R)^S\right) \ltimes \R^{2 S}. 
 \ee
 This group is maximal \cite{Ciambelli:2021vnn} (i-e any further local extension include the entire $\Diff(M)$ group) and universal \cite{Speranza:2017gxd} (i-e the same group is activated by higher derivative gravity theories).

 $\cS_\xi$ is the symplectic flux, which measures the failure of the Noether charge to be a Hamiltonian generator of the diffeomorphism symmetry.
 \be
\cS_\xi : =  \int_\Sigma \iota_\xi E  +
\int_S \iota_\xi \theta .
 \ee
We use here a different definition of the flux that the one in \cite{Freidel:2021cjp}. There  we considered the Hamiltonian flux denoted $\cF_\xi $ which was given by a pure boundary term and included the charge in its definition. The relation between the symplectic and Hamiltonian flux is  
 $\cS_\xi= \cF_\xi - Q_{\delta \xi} + \int_\Sigma \iota_\xi E$. The symplectic and Hamiltonian flux agree on-shell when the vector fields are field independent. Focusing on the symplectic flux is more adapted to our discussion. 
 It is important to note that when $\xi$ is field independent and tangent to the hypersurface $\Sigma$ then the flux is  a pure boundary term even when the  equations of motion are not imposed. This is relevant  for quantization, as this means that the tangential constraints $C_\xi$ form, in the bulk of $\Sigma$, a canonical representation of the diffeomorphism algebra.
 Note that this is not true for diffeomorphisms that move the surface $\Sigma$ transversally. In this case the flux is a  boundary term only on-shell.
 
 The name symplectic flux is justified by the fact that it controls how the symplectic potential transforms under diffeomorphism. If one denote $s_\xi:= \iota_\xi E +\rd \iota_\xi \theta$ the symplectic flux integrant we have that (see appendix \ref{proofA})
 \be\label{sympF}
 \cL_\xi \omega = s_\xi - s_{\delta \xi}. 
 \ee
 The integrated symplectic flux $\cS_\xi$ vanishes for the corner symmetry group  $G_S := \mathrm{Diff}(S)\ltimes \mathrm{GL}(2,\R)^S$, which means that  $G_S$  is represented canonically on the gravity phase space.  
On the other end, the normal subgroup  $ \R^{2S} $ of super-translations,  carries non zero flux and is not  represented canonically on the gravity phase space.
The presence of flux curtails our ability to understand the Noether charge $Q_\xi$ as an Hamiltonian generator of symmetry acting on the gravitational phase space.
Despite this, Barnich and Troeassert  \cite{Barnich:2009se,  Barnich:2010eb, Barnich:2011mi,Troessaert:2015nia} have proposed to consider a pseudo-Poisson bracket acting on the charges of boundary symmetry. This bracket has been extended in \cite{Freidel:2021cjp} to include all corner symmetries, not just the ones tangent to a given boundary, and  field dependency. This pseudo-Poisson bracket is only defined on the Noether charges and given by 
\be\label{pseudo-bra}
\{Q_\xi, Q_\psi \}:=  L_{\hat\xi} Q_\psi  - Q_{L_{\hat\psi}\xi}- I_{\hat\psi} \cS_\xi + \int_S \iota_\xi \iota_\psi L.
\ee
where $\cS_\xi$ is the symplectic flux. 
It was shown in \cite{Freidel:2021cjp} that when the Lagrangian is covariant this bracket forms a representation of the extended corner symmetry algebra. In other words we have
\be
 \{ Q_\xi, Q_\psi\}+Q_{\lbr\xi,\psi\rbr}=0.
\ee
Despite all these successes. The construction is unsatisfactory because it forbids us from arguing that the pseudo-bracket of charges descends from a canonical bracket defined on the gravity phase space. Without such an interpretation it is not possible to promote this bracket to a quantum commutator. This impossibility reflects the fundamental issues of open systems that carries non-trivial flux: One  do not expect them to be quantizable.
On the other end, the presence of a charge bracket satisfying Jacobi, suggests that it should be possible to overcome this difficulty and represent the charge action canonically. 
It turns out that the bracket which is quantizable is the modified pseudo-bracket
\be \label{pseudo-brap}
\{Q_\xi, Q_\psi \}'
:=\{Q_\xi, Q_\psi \}- Q_{L_{\hxi}\psi} + Q_{L_{\hat\psi}\xi}
= -  Q_{[\xi,\psi]_{\mathrm{Lie}}},
\ee
which provides a canonical representation of the diffeomorphism algebra instead of the algebroid.
It is useful  to express the pseudo-Poisson bracket \eqref{pseudo-brap} in terms of the symplectic structure.
Contracting \eqref{canonical} with $I_{\hat{\psi}}$ we see that we can  write this pseudo-bracket as
%\bea
%\Omega(\hxi,\hat\psi) =  L_{\hxi} Q_{\hat\psi} -Q_{L_{\hxi}\psi}
%-\I_{\hxi}\cS_{\psi}
%\eea
\bea
\{Q_\xi, Q_\psi \}' &=&  \Omega(\hxi,\hat\psi)  + \I_{\hxi}\cS_{\psi} - I_{\hat\psi} \cS_\xi + \int_S \iota_\xi \iota_\psi L,\cr
&=& \Omega(\hxi,\hat\psi) + \cL_\psi Q_\xi - I_{\hat\psi} \cS_\xi,\label{pseudobra}
\eea
where the second line follows from the equality 
$
\cL_\psi Q_\xi = \I_\hxi \cS_\psi + \int_S \iota_\xi \iota_\psi L,
$ proven in appendix \ref{proofB}.

\section{Extended Phase space}\label{EPS}
One of the main ingredient of the paper \cite{Donnelly:2016auv} is the introduction of an embedding field which extended the gravitational phase space and renders the covariant calculus covariant.
The embedding field also allows to cleanly distinguish the notion of gauge transformations from the notion of corner symmetries. This formalism was further developed by Speranza in \cite{Speranza:2017gxd} to include the case where the on-shell Lagrangian is non-vanishing. 

An embedding field is a map  $X: m \to M$, where $m$ is a reference manifold and $M$ is the spacetime. Given a gravity Lagrangian $L(g_{ab}, \phi)$ we can use this map to  define an extended Lagrangian ${L}^{\ext} = X^*(L)$
which depends on the metric $g$, the matter fields $\phi$, but also on the location of the spacetime region $X(m)$  supporting this Lagrangian. 
Embedding maps $X$  represent the local charts and allow us to locate the slice $\Sigma$ and its two sphere boundary $S=\pa \Sigma$ as fixed hypersurface and corner in the reference space\footnote{ Namely we can  assume that $m=\R^d$, that  $\Sigma$ is the image by $X$ of the hypersurface $\sigma=\{(0,\vec{x})| \vec{x}\in \R^{d-1}\}$ while $S$ is the image of the unit sphere $s$  inside $\sigma$.}.

The variation of the  extended Lagrangian gives 
\begin{equation}\label{dLext}
\delta L^{\ext} = X^*(\delta L + \mathcal{L}_{\chi} L) =X^*\left[  E+ \rd \left(\theta 
+  i_{\chi} L\right) \right].
\end{equation}
Here $\theta$ is the symplectic potential that would have been obtained for the original Lagrangian $L$, and hence depends only on the metric and field variations $(\delta g,\delta\phi)$ and not on $\delta X$.
$\chi $ denotes the Maurer-Cartan variational form
\be
\chi:= \delta X \circ X^{-1}. 
\ee
\subsection{Field space connection}

This variational form can be understood as a field space connection
valued into $\mathrm{diff}(M)$. The notion of field space connection was introduced by Gomes and Riello \cite{Gomes:2016mwl,Gomes:2018shn,Gomes:2018dxs} in the context of gauge theory. We generalize here this concept to the gravitational setting.
A field space connection is a vector valued one-form on Field space, i-e an element  $\co\in \Omega_1( \mathbb{F}, \mathfrak{aut}(P)) $, which satisfies
\be\label{axioms}
\I_\hxi \co = -\xi, \qquad \L_{\hxi} \co =  - \delta_\co \xi.
\ee
We introduce the covariant field space variation $\delta_\co$ which is a $(0,1)$ graded derivation\footnote{ There is a notational difference with \cite{Freidel:2021cjp}. Here we denote the field space Lie derivative along $\xi$ by $L_{\hxi}$ and the field space covariant variation by  $\delta_\co$. There we denoted the field space Lie derivative along $\xi$ by $\delta_{\xi}$ and didn't introduce a covariant variation.}
\be
\delta_\co : = \delta +\cL_\co. 
\ee
In components this means that 
$\co = \co^a \pa_a $ where $\co^a$ are variational one forms such that $I_\xi \co^a=-\xi^a$.\footnote{ The unusual sign follows from the fact that the lift map $\xi \to \hxi$ is an anti-morphism for field independent vector fields. In other words diffeomorphism action on field is a left action, i-e a morphism. While in the mathematical literature  one conventionally take the right action to be the gauge  action.
Since the right action is an anti-morphism, a minus sign is needed to compensate.}
Its curvature, given by 
\be
\curv[\co] : = \delta \co + \frac12 [\co,\co]_{\mathrm{Lie}}, 
\ee
is an horizontal form, and satisfies the Bianchi identity. This means that 
\be
\I_\hxi \curv [\co]=0, \qquad \delta_\co \curv =0.  
\ee
Both properties follow straightforwardly from the definition.
For instance
\be
L_\xi \co= \delta \I_\hxi\co +\I_\hxi \delta\chi
=- \delta \xi - \cL_\co \xi + \I_\hxi\curv[\co]
= -\delta_\co \xi  + \I_\hxi\curv[\co].
\ee
The curvature enters the commutator of covariant differential as an argument for the Lie  derivative
\be
\delta_\co^2 =\frac12 [\delta_\co,\delta_\co]
= \frac12[\delta,\delta] + [\delta, \cL_\co] + \frac12 [\cL_\co,\cL_\co]
= \cL_{\curv[\co]} .
\ee
The concept of an embedding field can now be formalized as a choice of a field-space connection
which is \emph{flat}. This flat connection is the Maurer-Cartan connection for the embedding field.
Given any form $\alpha$ we have 
\be
\delta ( X^* \alpha) = X^*(\delta_\co \alpha). 
\ee
With this understanding we do not have to restrict to global embedding fields.\footnote{ 
Another class of field space connection arises through gauge fixing. If $G(g)=0$ is a gauge fixing condition then one can associate to this a field space connection $\chi_G$ which is such that 
$\delta_{\chi_G} G=0$. }

Given a connection $\co$ we can define the covariant Lie derivative  by 
\bea\label{covL}
L^\co_\xi &:=&[\delta_\co, I_\hxi]= [\delta, I_\hxi]+[\cL_\co, I_\hxi] = \L_\hxi -{\cL_\xi}.
\eea
while the anomaly operator and the bracket can be expressed as the commutators
\be
\Delta_\xi = -[\I_\hxi,L_{\hco}- \cL_{\co}], \qquad [L^\co_{\hxi}, \I_{\hat\psi} ] = I_{\widehat{\lbr \xi , \co\rbr}}.
\ee
This is shown in appendix \ref{proofC}. Finally, an important property of the field-space connection is that it is anomaly free
\be
\Delta_\xi \co = L_{\hxi}\co - \cL_\xi \co - I_{\delta \hxi} \co 
= -\delta_\co \xi + \cL_\co \xi +\delta\xi=0. 
\ee

\subsection{Extended symplectic structure}
From the variation \ref{dLext} we conclude that the  extended symplectic potential is given by $\theta^{\ext} = X^* \theta_\chi $ where $\theta_\chi$ is the covariant symplectic potential  
\be
\theta_\chi:= \theta + \iota_\chi L,
\ee
and  $\chi$ is the flat Maurer-Cartan connection. This potential is covariant when $\theta$ is, i-e we have $\Delta_\xi \theta_\co=0$.
The Noether current  \eqref{Ncur} is simply given by the contraction of the covariant potential along the Lie derivative 
\be
\I_{\hxi}\theta_{\co} = J_\xi= C_\xi +\rd q_\xi. 
\ee
where $q_\xi$ is the corner charge aspect introduced earlier. 
From this we define   the covariant symplectic structure:
\be
\omega_\co:= \delta_\co \theta_\co,
% -  \iota_{\curv[\co]}L, 
\qquad 
\Omega_\chi: =\int_\Sigma \omega_\chi.
\ee
In appendix \ref{proofD} the covariant  symplectic form is expanded  in terms of components. The expansion shows that, although the modification of the symplectic potential is a bulk modification, the covariant symplectic structure differs  from the non-covariant one by an on-shell term plus a corner term:
\be\label{mocha}
 \omega_\co =  \omega +  \iota_\co E
+ \rd\left(  \iota_{\co}\theta + \tfrac12 \iota_\co \iota_\co L \right)
.
\ee
This symplectic structure (which already appears in \cite{Speranza:2017gxd}) is covariantly closed $\delta_\co\omega_\co=0$ and satisfies the crucial property of being conserved on-shell $
\rd \omega_\co= \delta_\co E.$

Integrating \ref{mocha} on a slice $\Sigma$  leads to the covariant symplectic  form
\be
\boxed{\,\, \Omega_\chi= \Omega + \cS_\chi + \frac12 \int_S \iota_\chi \iota_\chi L \,\,}
\ee
One can now establish our first central result: The action of  field independent diffeomorphism $\delta \xi=0$ is canonical. More precisely, one has that 
\be 
-I_{\hat\xi} \Omega_\co = \delta_\chi Q_\xi - Q_{\delta \xi}.
\ee 
This follows from the fundamental canonical relation \eqref{canonical} and the identity 
\be\label{transid}
I_{\hat \xi} \cS_{\co}- \int_S \iota_\xi \iota_\chi L= -\cS_{\xi} -\cL_{\co} Q_\xi,
\ee proven in appendix \ref{proofC}. The introduction of the connection into the symplectic structure has allowed us to reabsorb the flux term into the symplectic potential.
This means that the diffeomorphism action is Hamiltonian (even off-shell!) and that the bracket of charge is therefore given by the usual Poisson bracket  for the extended phase space. 
\be
\{ Q_\xi, Q_\psi \}_\co = \Omega_\co( \hat{\xi},\hat{\psi}) = L_\hxi^\co Q_\psi -  Q_{L_{\hat\xi} \psi}.
%=   Q_{\lbr \xi, \psi\rbr}.
\ee
In the second equality we used \eqref{covL}.
Our second main result is the fact that this canonical Poisson bracket coincides with the  pseudo Poisson Bracket described in \eqref{pseudo-brap}. This follows from
\bea
 \Omega_\co( \hat{\xi},\hat{\psi})= I_{\hat\psi} I_{\hat\xi}\Omega_\chi &=& I_{\hat\psi} I_{\hat\xi}\Omega + 
 I_{\hat\psi} \left(I_{\hat\xi}\cS_\chi -  \int_S \iota_\xi \iota_\co L\right)\cr
 &=&\Omega ( \hat{\xi},\hat{\psi}) - I_{\hat\psi} \cS_{\xi} + \cL_{\psi} Q_\xi= \{Q_\xi, Q_\psi \}' .
 %\cr
% &=&-L_{\hat \psi} Q_\xi + \cL_{\psi} Q_\xi
\eea
where we used \eqref{transid} in the last equality. And we recognize the expression of the pseudo-bracket given in \eqref{pseudobra}. This equality between the extended canonical bracket and the pseudo-bracket ensures that 
\be
\{ Q_\xi, Q_\psi \}_\co = - Q_{[\xi,\psi]_{\mathrm{Lie}}}.
\ee
The covariant calculus that we have develop allows us to give a simple proof of the fact that 
\be
L_\hxi^\co Q_\psi -  Q_{L_{\hat\xi} \psi}=- Q_{[\xi,\psi]_{\mathrm{Lie}}}.
\ee First, from the vanishing anomaly condition $\Delta_\xi \theta_\co=0$ we get that 
$L^\co_\xi \theta_\co = I_{\delta \xi} \theta_\co=C_{\delta \xi} + \rd q_{\delta \xi} $ and therefore
\bea
L_\hxi^\co Q_\psi &=&  \int_{\Sigma} L_\hxi^\co I_{\hat\psi} \theta 
=  \int_{\Sigma}  I_{\hat\psi} L_\hxi^\co \theta_\co + \int_{\Sigma}  [L_\hxi^\co, I_{\hat\psi}] \theta_\co \cr
& = &I_{\hat\psi} Q_{\delta \xi}
-\int_{\Sigma}  I_{\widehat{\lbr \xi,\psi\rbr}} \theta_\co = -Q_{\lbr \xi,\psi\rbr } + Q_{L_{\hat\psi} \xi}\cr
& =& -Q_{[\xi,\psi]_{\mathrm{Lie}}} + Q_{L_{\hat \xi} \psi}. 
\eea
\section{Conclusion}
In this note, we have established that the extended corner symmetry algebra can be represented canonically on the gravitational phase space enlarged by the presence of the embedding field.

This result is presented in a technical manner and clearly begs for a more profound conceptual explanation that we expect to develop in the future.  What is puzzling is that we are talking about an open system, which can have degrees of freedom leaving and entering the system through its boundary. We don't expect an Hamiltonian action.  In particular, the dressed symplectic form is not conserved under super-translations. Nevertheless, the symmetry charges are Hamiltonian. At a colloquial level, this phenomenon is simply that the embedding field acts as an extremely thorough gatekeeper who faithfully records what symmetry charge information leaves or enters the system. 

In \cite{Donnelly:2016auv} we showed that the introduction of the embedding field creates two type of transformations which commute with one another. The left action, given by 
$
L_{\hxi} g_{ab}= \cL_\xi g_{ab},
$ and $ L_{\hxi} X = \xi ,$
 is the action we have studied here and represents  infinitesimal automorphisms. 
 There is also the right action. It is labeled by vector fields $v$ on the source $m$ of the embedding map $X: m\to M$. This action is given by 
$
L_{\check{v}} g_{ab}= 0,$ and $L_{\check{v}} X = \rd X (v). $
It represents the  active translation of the corners. The canonical study of such transformations is awaiting.

 Note that in \cite{Donnelly:2016auv} we used the embedding field for a different strategy than the one here. There it was used to shift the symplectic potential and defined 
 $\theta^{\mathrm{ext}} =\theta_\co+ \rd q_\co$, introducing edge modes \cite{Freidel:2020xyx,Freidel:2020svx,Freidel:2020ayo}. Such a transformation renders the left action pure gauge and promote the right translations as symmetries.
 It would also be interesting to see if we can also implement this strategy here through the choice of a boundary Lagrangian $\ell$ and shift 
 the gravity  Lagrangian by a boundary term $L \to L+ \rd \ell$.
 What is required is that $q_\co$ appears as the corner symplectic potential of $\ell$.
 It is natural to wonder if the canonical representation of symmetry charges survives the introduction of edge modes.

\section*{Acknowledgments}
I would like to thank D. Pranzetti for his precious help and support. I would like to also thank C. Zwikel for discussion and feedback and L. Ciambelli and R. Leigh for sharing an early version of their draft.
Research at Perimeter Institute is supported in part by the Government of Canada through the Department of Innovation, Science and Economic Development Canada and by the Province of Ontario through the Ministry of Colleges and Universities.

\appendix\label{proofs}
\section{Fundamental canonical relation and flux}\label{proofA}
The validity of the fundamental canonical relation \eqref{Fundr} straightforwardly follows from the covariance condition  $\Delta_\xi \theta=( L_{\hxi} - \cL_{\xi} -\I_{\delta \hxi})\theta =0$ and after evaluating 
\bea
L_{\hxi}\theta &=& \I_\hxi \omega + \delta \I_{\hxi} \theta = \I_\hxi \omega +\delta \left( C_\xi +\rd q_\xi + \iota_\xi L \right),\cr
\cL_{\xi} \theta &=& \iota_\xi \rd \theta + \rd \iota_\xi \theta = \iota_\xi \delta L + 
( \iota_\xi E+ \rd \iota_\xi \theta ) , \cr
\I_{\delta \hxi}\theta &=& \iota_{\delta \xi} \delta L  + C_{\delta \xi} +\rd q_{\delta \xi}.
\eea
To establish the local  symplectic flux law  \eqref{sympF} we use that
\bea
\cL_\xi \theta &=& \rd \iota_\xi \theta + \iota_\xi(E+\delta L) = s_\xi + \iota_\xi \delta L,\cr
\cL_\xi \omega &=& \delta \cL_\xi \theta - \cL_{\delta \xi}\theta = \delta s_\xi - s_{\delta \xi} + \delta (\iota_\xi \delta L) - \iota_{\delta \xi} \delta L = \delta s_\xi - s_{\delta \xi} .
\eea
where we have introduced $s_\psi := \rd \iota_\psi \theta +  \iota_\psi E$ the symplectic flux integrand:
$\cS_\psi =\int_\Sigma s_\psi$.
\section{Flux Charge relation}\label{proofB}
We now prove the identity $\cL_\psi Q_\xi = \I_\xi \cS_\psi + \int_S \iota_\xi \iota_\psi L$ used in section \ref{CovPS}.
We work at the level of charge aspects
\bea
\cL_\psi q_\xi &=& \cL_\psi (\I_{\hxi} \theta -\iota_{\xi} L) 
\cr
&=& \I_{\hxi} (\rd \iota_\psi \theta + \iota_\psi \rd \theta) - \cL_\psi \iota_{\xi} L \cr
&=&  \I_{\hxi} (\rd \iota_\psi \theta +  \iota_\psi E+ \iota_\psi \delta L ) - \cL_\psi \iota_{\xi} L \cr
&=& \I_{\hxi} s_\psi+ \iota_\psi \cL_{\xi} L - \cL_\psi \iota_{\xi} L,
\eea
 To conclude we use that 
\be 
\iota_\psi \cL_{\xi} L - \cL_\psi \iota_{\xi} L =
\iota_\psi \iota_\xi \rd  L - \rd \iota_\psi \iota_{\xi} L = \rd  \iota_{\xi} \iota_\psi L,
\ee
and we integrate over $\Sigma$.
We now evaluate
\bea
\I_\hxi s_\psi +  \rd \iota_\xi \iota_\psi L &=&
\I_\hxi \iota_\psi  E + \I_{\hxi} \rd \iota_\psi \theta +  \rd \iota_\xi \iota_\psi L \cr
&=& \iota_\psi  \rd C_\xi  +  \rd \iota_\psi (\I_{\hxi}\theta -  \iota_\xi  L)\cr
&=& (\iota_\psi  \rd + \rd \iota_\psi  )( C_\xi +\rd q_\xi).
\eea
which gives the identity 
\be
\cL_{\psi} Q_\xi =  \I_\hxi S_\psi +  \int_S \iota_\xi \iota_\psi L,
\ee
after integration over $\Sigma$.

\section{Commutators}\label{proofC}

The variational Cartan calculus 
The operator $I_\hco$ satisfies the following properties
\bea
[I_\hco, I_{\hxi}]&=& -I_{I_\hxi \hco} =I_{\hxi}, 
\qquad 
[I_\hco, \delta]:= L_\hco,\qquad [I_{\hco}, \cL_{\co}]= - \cL_{\co}
%\cr[I_\hxi, I_{\delta \hco}]&=& I_{I_\hxi \delta \hco} =I_{\cL_{\hco} \hxi},\cr
\eea
From this we get that $  [I_{\hco}, \delta_\co] = L_{\hco}- \cL_{\co}$.
\bea
[I_\hco, L_{\hxi}]&=&  I_{\widehat{\lbr \xi , \co\rbr}}= I_{\delta \xi} + I_{L_{\hco}\xi}, \cr
\lbr \xi , \co\rbr&=& 
{[\xi, \co]_{\mathrm{Lie}} +  L_{\hco} \xi -L_\hxi \co}
= \delta_\co \xi + (L_{\hco}-\cL_{\co}) \xi.
\eea
From this we establish that 
  \bea
  [I_\hxi, L_{\hco}]&=& [I_\hxi, [I_{\hco}, \delta]]= [[I_\hxi, I_{\hco}], \delta]
+ [I_\hco, [I_{\hxi}, \delta]]= -[I_{\hxi}, \delta] + [I_\hco, L_{\hxi}]\cr
&=& - L_{\hxi} +I_{\widehat{\lbr \xi , \co\rbr}}.
\eea
This means that 
we get
\bea
[I_{\hxi}, \delta_\co] &=& [I_{\hxi}, \delta] + [I_{\hxi }, \cL_{\co}]= L_\hxi-\cL_\xi,\cr
 [I_{\hco}, \delta_\co]&=&   [I_{\hco}, \delta] + [I_{\hco}, \cL_{\co}]=L_{\hco}- \cL_{\co},\cr
 [\I_\hxi,L_{\hco}- \cL_{\co}] &=& - (L_\hxi-\cL_{\xi} - I_{\delta_\co \xi})
 = - \Delta_{\xi}
  \eea
  We also have 
\bea
[I_{\hxi}, \delta_\co] &=& [I_{\hxi}, \delta] + [I_{\hxi }, \cL_{\co}]= L_\hxi-\cL_\xi,\cr
 [I_{\hco}, \delta_\co]&=&   [I_{\hco}, \delta] + [I_{\hco}, \cL_{\co}]=L_{\hco}- \cL_{\co},\cr
 [\I_\hxi,L_{\hco}- \cL_{\co}] &=& - (L_\hxi-\cL_{\xi} - I_{\delta_\co \xi})
 = - \Delta_{\xi}.
  \eea
%  from which we conclude that 
%  \bea
% [L_{\hco}- \cL_{\co},\delta_\co] &=& [[I_{\hco}, \delta_\co],\delta_{\co}]= 
%  \tfrac12 [I_{\hco}, [\delta_\co,\delta_{\co}]] 
%  =[I_{\hco},\cL_{\curv}]= \cL_{I_\xi \curv}=0.
%  \eea
%While 
%\bea
%[\I_{\delta \hco}, \Delta_\xi]&=&[\I_{\delta \hco} , L_\hxi] - [\I_{\delta \hco}, \cL_\xi]- [\I_{\delta \hco}, I_{\delta\xi}]\cr
%&=& - L_{\delta L_\hxi \hco}
%\eea
\section{Charge algebra}\label{proofD}

One first establish the key identity \cite{Speranza:2017gxd}
\be\label{warmup}
 \cL_\co \iota_\co = \frac12 \left( 
 \iota_{[\co,\co]} 
+ \rd \iota_\co \iota_\co- \iota_\co \iota_\co \rd 
\right)
\ee
which follows form the Cartan identity 
$[\cL_\xi,\iota_{\rho}]=\iota_{[\xi,\rho]}$
and the fact that $\cL_{\xi}$ is a $(0,1)$ and $\iota_\xi$ is a $(-1,1)$ graded differential operator 
\bea
\cL_\co \iota_\co 
&=& \iota_{[\co,\co]} - \iota_\co \cL_\co\cr
&=&\iota_{[\co,\co]} -  \iota_\co\rd \iota_\co -  \iota_\co \iota_\co \rd\cr
&=&\iota_{[\co,\co]} -  ( \rd \iota_\co \iota_\co +\iota_\co\rd \iota_\co) 
+ \rd \iota_\co \iota_\co-  \iota_\co \iota_\co \rd\cr
&=&\iota_{[\co,\co]} -  \cL_\co \iota_\co 
+ \rd \iota_\co \iota_\co-  \iota_\co \iota_\co \rd.
\eea
From which we get \eqref{warmup}.

We can now use it to evaluate the covariant symplectic structure
\bea
\omega_\co 
&=& (\delta +\cL_{\co})(\theta + \iota_\co L)  \cr
&=& \delta \theta + \delta \iota_\co L +\cL_{\co}\theta +\cL_\co \iota_\co L  \cr
&=& \omega + \iota_{\delta \co}L -\iota_\co \delta L 
+ \cL_{\co}\theta + \frac12 \left( 
 \iota_{[\co,\co]} 
+ \rd \iota_\co \iota_\co-  \iota_\co \iota_\co \rd 
\right)L  ,\cr
&=& \omega  -\iota_\co \rd \theta + \iota_\co E
+ \cL_{\co}\theta + \tfrac12  \rd \iota_\co \iota_\co L +\iota_{\curv[\co]} L\cr
 &=&
 \omega +  \iota_\co E
+ \rd\left(  \iota_{\co}\theta + \tfrac12 \iota_\co \iota_\co L \right)
 \eea
 where we used \ref{warmup} in the third line and ${\curv[\co]} =0$ in the last equality.
%\bea
%\omega_\co 
%&=& (\delta +\cL_{\co} + \I_{\hcF})(\theta + \iota_\co L) -\iota_{\curv[\co]} L \cr
%&=& \delta \theta + \delta \iota_\co L +\cL_{\co}\theta +\cL_\co \iota_\co L {\color{blue}-\iota_{\curv[\co]} L }\cr
%&=& \omega + \iota_{\delta \co}L -\iota_\co \delta L 
%+ \cL_{\co}\theta + \frac12 \left( 
% \iota_{[\co,\co]} 
%+ \rd \iota_\co \iota_\co-  \iota_\co \iota_\co \rd 
%\right)L  -\iota_\curv L,\cr
%&=& \omega  -\iota_\co \rd \theta + \iota_\co E
%+ \cL_{\co}\theta + \tfrac12  \rd \iota_\co \iota_\co L\cr
% &=&
% \omega +  \iota_\co E
%+ \rd\left(  \iota_{\co}\theta + \tfrac12 \iota_\co \iota_\co L \right)
% \eea
We now focus on the proof of \eqref{transid}. We denote $s_\co= \iota_\co E +\rd \iota_\co \theta$ the symplectic flux integrand and we evaluate

\bea
I_\hxi  s_\co + \tfrac12 I_{\hxi} \rd(\iota_\co \iota_\co L) &=&I_\hxi  \iota_\co E
+ \rd I_\hxi \left( \iota_{\co}\theta + \tfrac12 \iota_\co \iota_\co L \right) \cr
 &=& - ( \iota_\xi E  + \iota_\co \rd C_\xi )
- \rd \left( \iota_{\xi}\theta + \iota_\co \left(I_\hxi \theta - \iota_{\xi}  L \right) \right) \cr
&=& - ( \iota_\xi E  + \rd \iota_{\xi}\theta + \cL_{\co}( C_\xi  + \rd q_\xi ) ).
\eea
We obtain \eqref{transid} after integration.
We can finally evaluate \eqref{mocha}
\bea
I_\hxi \omega_\co &=&I_\hxi\omega 
+I_\hxi  \iota_\co E
+ \rd I_\hxi \left( \iota_{\co}\theta + \tfrac12 \iota_\co \iota_\co L \right)\cr
&=& - \delta \left(  C_\xi + \rd q_\xi  \right) + \left[\iota_\xi E+ C_{\delta \xi} +
\rd (\iota_\xi \theta + q_{\delta \xi}) \right]
- ( \iota_\xi E  + \iota_\co \rd C_\xi )
+ \rd I_\hxi \left( \iota_{\co}\theta + \tfrac12 \iota_\co \iota_\co L \right)\cr
&=& - \delta \left(  C_\xi + \rd q_\xi  \right) +
\rd (\iota_\xi \theta ) 
-  \iota_\co \rd C_\xi 
- \rd  \left( \iota_{\xi}\theta  +\iota_\co I_\hxi \theta+ \iota_\xi \iota_\co L \right)
+  C_{\delta \xi} +\rd q_{\delta \xi} \cr
&=& - \delta \left(  C_\xi + \rd q_\xi  \right)  
-  \iota_\co \rd C_\xi 
- \rd  \left( \iota_\co C_\xi +\iota_\co \rd q_\xi \right)+  C_{\delta \xi} +\rd q_{\delta \xi} \cr
&=& - \delta_\co \left(  C_\xi + \rd q_\xi  \right) + C_{\delta \xi} +\rd q_{\delta \xi}.
\eea
where we have used the on-shell definition of the Noether charge
\be
C_\xi+ \rd q_\xi= I_\hxi  \theta -\iota_\xi L\,.
\ee

%From the relation above, assuming $\delta \xi=0$ one can prove 
% \bea
% I_\hxi \rd( \delta q_\co +\iota_\co \rd q_\co) &=& 
% \rd(( L_\hxi - \cL_\xi )q_\co +\delta_\co q_\xi )  
% = \rd(q_{L_\hxi\co} - q_{\cL_\hxi\co} +\delta_\co q_\xi )\cr
% &=& \rd(-q_{\delta_\co \xi } - q_{\cL_\hxi\co} +\delta_\co q_\xi )\cr
% &=& - \rd (q_{\delta \xi }) + \delta_\co (\rd q_\xi )
% \eea

%\bibliographystyle{utphys}
\bibliographystyle{bib-style2}
\bibliography{biblio-fluxes.bib}

\end{document}